\documentclass[12pt,aps,nofootinbib,showkeys]{revtex4}
\usepackage{color}
\usepackage{multirow}
\usepackage{amsmath}
\usepackage{esvect}
\usepackage{soul}
\usepackage[normalem]{ulem}
\usepackage{tikz}
\usetikzlibrary{angles,quotes}

\usepackage{amsfonts}    
\usepackage{amssymb}
\usepackage{latexsym}
\usepackage{graphicx}
\usepackage{physics}

\usepackage{booktabs}
\usepackage[english]{babel}

\begin{document} 


\title{The Weakly Bound States in Gaussian Wells: From  the Binding Energy of Deuteron to the Electronic Structure of Quantum Dots
}





 \author{G. Rodriguez-Espejo}
 \affiliation{Facultad de F\'isica, Universidad Veracruzana, C. P. 91097, Xalapa, M\'exico.}

  \author{J. A. Segura-Landa}
 \affiliation{Facultad de F\'isica, Universidad Veracruzana, C. P. 91097, Xalapa, M\'exico.}

  \author{J. Ortiz-Monfil}
 \affiliation{Facultad de F\'isica, Universidad Veracruzana, C. P. 91097, Xalapa, M\'exico.}

 \author{D~.~J.~ Nader}
 \email{daniel.juliannader@upol.cz} 
\affiliation{Department of Optics, Faculty of Science, Palack\'y University, Olomouc 77146, Czech Republic}



\begin{abstract}

Gaussian potentials serve as a valuable tool for the comprehensive modeling of short-range interactions, spanning applications from nuclear physics to the artificial confinement of electrons within quantum dots. This study focuses on examining the lowest states within Gaussian wells, with particular emphasis on the weakly bound regime. The analysis delves into the asymptotic behavior of the exact wave function at both small and large distances, motivating the development of a few-parametric Ansatz which is locally accurate and yields to a fast convergent basis set. To validate its efficacy, we assess its convergence rate using a toy model of Nuclear Physics, specifically for Deuteron. Furthermore, we employ the expansion of the energy close to the threshold to derive an analytical formula for the binding energy of the deuteron whose accuracy improves as the effective parameter approaches the critical. In concluding our investigation, we evaluate the performance of our Ansatz as an orbital in the exploration of the electronic structure of a two-electron quantum dot.
\end{abstract}


\maketitle

\section{\label{sec:intro}Introduction}

One of the simplest non-analytically solvable short-range potentials is represented by the Gaussian well \cite{FernandezGaussianWell, NandiGaussianWells}, expressed as

\begin{equation}
\label{GaussV}
V(r)=- V_0,e^{-\lambda r^2},,
\end{equation}

where $V_0$ and $\lambda$ are positive parameters regulating the depth and width of the potential, respectively. The radial coordinate in $d$-dimensional space is denoted as $r$. It is important to note that the Gaussian well is commonly employed for investigating resonances in various quantum systems, including $^{12}\text{C}+p$, $\alpha+{}^3\text{He}$, and $\alpha+\alpha$, as documented in \cite{Descouvemont_2010}. Furthermore, in two and three dimensions, Gaussian wells and barriers have proven to be effective in the examination of diverse quantum phenomena, such as quantum rings and polarons \cite{Xiao_2019, Gylfadottir2006, Cuevas_2013}. 

Specifically, quantum wells are often employed to simulate the artificial confinement of electrons within quantum dots \cite{Chaudhuri,PhysRevB.62.4234,Boyacioglu_2007,FewelectronsemiconductorquantumdotswithGaussianconfinement,XIE20082828}, as opposed to harmonic confinement \cite{Poszwa2016,Nader2017}, which does not allow for dissociation. Additionally, the Gaussian potential offers a straightforward approach to modeling short-range interactions among nucleons \cite{Barnea2015}.

    Due to the exponential decay inherent in the Gaussian function, Gaussian wells in any dimension contain a finite number of bound states, thereby giving rise to critical parameters at which the wave function loses its square normalizability. The determination of these critical parameters has posed a formidable challenge for approximation methods in quantum mechanics across various shapes of confining potentials. This challenge is particularly explored for screened Coulomb potentials, such as the Yukawa potential, as evidenced by prior investigations \cite{https://doi.org/10.1002/qua.25108,delValle2018,PhysRevA.50.228,10.1093/ptep/ptx107,PhysRevA.48.220,Diaz_1991,doi:10.1021/jp9820572}, as well as for systems involving critical Coulombic charges \cite{Sergeev,PhysRevA.84.064501,PhysRevA.97.022503,PhysRevA.105.052806,https://doi.org/10.1002/qua.26879}.

Within the broader context of studies concerning critical parameters associated with central potentials in three dimensions, a limited number of estimations for critical parameters pertaining to Gaussian wells can be found in the literature \cite{Liverts_2011,FERNANDEZ2013580}. 

 n the particular scenario of Gaussian wells, diverse methodologies have been utilized to achieve precise estimations of critical parameters. A common technique involves setting the eigenvalue of the Hamiltonian to a threshold value, typically zero in this context, and subsequently numerically solving the associated differential equation to ascertain the critical parameter \cite{Liverts_2011}. Despite the high accuracy offered by this method in estimating critical parameters, it lacks the capability to provide insights into the energy and wave function characteristics within the weakly bound regime, marked by parameters closely approaching their critical values.
 
 In this manuscript, we address the radial Schr"odinger equation's solution for a particle under Gaussian confinement in both two and three dimensions. Utilizing the Lagrange Mesh Method (LMM), we obtain a highly accurate approximation of the energy and subsequently estimate the critical parameters governing the system. We further investigate the energy behavior for the lowest states near the threshold, employing an expansion proposed in \cite{KLAUS1980251} based on spectral theory.

Next, we introduce a compact wave function with three variational parameters, demonstrating local accuracy. Constructed from asymptotic limits at small and large distances, this wave function reproduces up to eight significant digits of energy compared to LMM results, particularly when sufficiently distant from the critical value. Applying this wave function to a nuclear physics toy model, specifically for deuteron, the superposition of different configurations using our Ansatz exhibits a rapid convergence rate of energy concerning the number of terms in the superposition.

Finally, we employ the locally accurate compact Ansatz as an orbital to construct a wave function for a two-electron quantum dot, where the one-body potential is modeled by a Gaussian well. The obtained results prove highly competitive when compared to previous literature findings.  

\section{Generalities}

The Hamiltonian of a $d$-dimensional particle trapped in a radial Gaussian well is 
\begin{equation}
\label{HamOriginal}
\hat{H}=-\frac{\hbar^2}{2m}\nabla_d^2\ -\ V_0\,e^{-\lambda r^2} \ , \qquad r\in[0,\infty)\ ,
\end{equation}
where $\nabla^2_d$ is the $d$-dimensional Laplacian operator and $r$ is the distance from the center of coordinates (radial coordinate). 
Since the potential is central, the angular part can be easily separated from the radial equation
using $d$-dimensional spherical coordinates, see \cite{Harmonics}. In particular, we are interested in $d=2,3$. After separation of variables, the radial Schr\"odinger equation in $d$-dimensions reads
\begin{equation}
-\frac{\hbar^2}{2m}\left(\frac{d^2R(r)}{dr^2}\ +\ \frac{d-1}{r}\frac{d R(r)}{dr} \right)\ +\  \left(-V_0e^{-\lambda r^2}\ +\ \frac{\ell(\ell+d-2)}{2\,r^2}\right)R(r)\ =\ ER(r)\ ,   
\end{equation}
where $\ell=0,1,2,...,$ is the angular quantum number. By using the factorization $R(r)=r^{-\frac{d-1}{2}}\tilde{R}(r)$, we remove the term that contains the derivative of first order, leading to the reduced radial Schr\"odinger equation
\begin{equation}
\label{radialSeq}
-\frac{\hbar^2}{2m} \frac{d^2 \tilde{R}(r)}{dr^2} 
 +\left(-V_0 e^{- \lambda r^2}+
\frac{(d+2 \ell-1) (d+2 \ell-3)}{8r^2}\right)\tilde{R}(r) =E\tilde{R}(r)\,.
\end{equation}
 In two dimensions the angular part of the total wave function is given by $e^{-il\phi}$ where $\phi$ is the angular polar coordinate. In the three dimensional case, the angular part is given by spherical harmonics $Y_{lm}(\theta,\phi)$, being eigenfunctions of the angular momentum operator $\hat{L}^2$ with the eigenvalues $\hbar^2l(l+1)$ . Here, $\theta$ and $\phi$ are the angular spherical coordinates. Thus, the problem is reduced to solve the  equation (\ref{radialSeq}).
For the one-dimensional case, it was shown in  \cite{FernandezGaussianWell} that, introducing the scaling coordinate $x \to \sqrt{\lambda}x^\prime$ into (\ref{radialSeq}), the energy transforms as 
\begin{equation}
    E(V_0,\lambda,\frac{\hbar^2}{m})= \frac{\hbar^2}{m}\lambda E(\frac{mV_0}{\lambda\hbar^2},1,1).
    \label{transform}
\end{equation}
Thus, the parameter $\lambda$ can be set to the unity and the energy only depends on the ratio $v_0=\frac{mV_0}{\lambda\hbar^2}$, which is the effective parameter of the potential. It can be easily demonstrated that the energy relation shown in (\ref{transform}) holds for arbitrary $d$.

Regarding the determination of the number of bound states for a given depth and width of the Gaussian well. It is established that specific critical values of $V_0$ and $\alpha$ exist, beyond which a particular state ceases to be bound and transitions into the continuum regime. Any confining one-dimensional potential well which satisfies\footnote{Weaker conditions may be imposed on $V(r)$ to guarantee the existence of at least one bound state, see e.g. \cite{Brownstein} and references therein.} $V(r)\leq0$ and $\displaystyle{\lim_{|r| \to \infty}} V(r) = 0$, contains at least one bound state independently on the  depth or width of the potential. Under analogous considerations, it can be demonstrated that any two-dimensional spherically symmetric confining potential contains at least one bound state \cite{Yang}.
Contrarily, in the case of radial potentials in three dimensions, it is established that no bound states exist if the confining potential lacks adequate width or depth \cite{Kocher1977}. 
For potentials that satisfy the condition $\displaystyle{\lim_{|r| \to \infty}} V(r) = 0$, their critical parameters correspond to the particular values where the energy of such state vanishes
 $$E(V_0^c,\lambda^c)=0\,.$$
 Hence, as the parameters $V_0$ and $\lambda$ approach their critical values, represented by $V_0^c$ and $\lambda^c$, the state undergoes a transition into a weakly bound regime. In this context, the wave function tends to lose its square normalizability and becomes notably diffuse in space. Under this circumstances, the estimation of any physically relevant quantity is not trivial.



\subsection{Energy in the vicinity of the critical parameter}

In three dimensions, the energy of $s$-states ($l=0$) close to the threshold \cite{KLAUS1980251} behaves as 
\begin{equation}
 \label{expansiona}
 E_{n,l=0}(v_0)=\sum^\infty_{n=2} \gamma^{(n,l)}_n(v_0-v_0^c)^n\,,
\end{equation}
while the energy of the states with an angular momentum $l\neq0$  behaves as
\begin{equation}
 \label{expansionb}
 E_{n,l}(v_0)=\sum^\infty_{n=2} \xi^{(n,l)}_n(v_0-v_0^c)^{n/2}\,.
\end{equation}

The coefficients $a_n$ and $b_n$ can be obtained directly by interpolating  the numerical results for the energy close to the threshold. In three dimensions for the case of states with non-zero angular momentum, the coefficient $b_2$ of the leading term can be obtained also by means of the Hellman-Feynman theorem via expectation values 
\begin{equation}
 \label{HellmanFeynman}
 \xi_2^{(n,l)}=\frac{\partial E_{n,l}}{\partial v_0}=\langle \Psi_{n,l}| \frac{\partial \hat{H}}{\partial v_0} |\Psi_{n,l} \rangle =- \langle \Psi_{n,l}   | e^{-r^2} | \Psi_{n,l} \rangle \,,
\end{equation}
using (\ref{transition}). For the case of two dimensions, only the leading term is known for the lowest states
\begin{eqnarray}
\label{C2d}
E_{n=2,l=0}(v_0)&=&\eta_1e^{\eta_2/v_0-v_0^c}\, \nonumber\\
E_{n=1,l=1}(v_0)&=&\eta_1(v_0-v_0^c)/\log(v_0-v_0^c)\, \nonumber \\
E_{n=1,l=2}(v_0)&=&\eta_1(v_0-v_0^c)\,.
\end{eqnarray}

\subsection{Connection between different dimensions and angular momenta }

There is an interesting connection between the spectrum of radial Schr\"odinger operators in different dimensions and different angular momentum. For unknown reasons to the present authors, the straightforward relation is not frequently discussed. The connection can be explicitly seen and established in (\ref{radialSeq}) by noting  that for given ($d,\ell$) and ($d',\ell'$) that satisfy the equation
\begin{equation}
  d\ +\ 2\ell\ =\ d'\ +\ 2\ell' ,  
  \label{connection}
\end{equation}
the spectral problem defined by equation (\ref{radialSeq}), together with normalizability conditions for eigenfunctions, remains unchanged.  As a consequence, the spectra are also related. For example, the three-dimensional  ($d=3$) spectrum with angular momentum $\ell=0$ ($p$-states) coincides  with the one-dimensional (antisymmetric states\footnote{At $d=1$ an even extension of  the potential is needed. Furthermore,  the quantum number $\ell$ plays the role of the parity quantum number, usually denoted by $p=0,1$.}) $d'=1$ with $\ell'=1$. In general, from (\ref{connection}) it is easy to prove that a pair of different dimensions is \textit{connected} if $d+d'\,(\text{mod}\,2)=0$.
Since connection between the spectral problems is at the level of energies and wave functions, critical screening parameters are connected as well. 

\section{The Methods }

\subsection{The Lagrange Mesh Method}

The Lagrange Mesh Method (LMM) is a very efficient numerical method to solve the Schr\"odinger equation via exact diagonalization\cite{BAYEPhysRep,Baye_2011}. The method consist in finding a matrix representation for both the kinetic and potential term of the Hamiltonian in the Lagrange basis (see Appendix \ref{Appendix A}). In this work, we made very similar considerations as in \cite{delvalle2023two}. In the two-dimensional case we use the generalized Laguerre mesh for the ground state. For all the bounded states in two dimensions as well as all the bounded states in three dimensions we use the Laguerre mesh.  

\subsection{The Variational Method}

In the framework of the variational method, we designed a compact trial wave function by matching the asymptotic expansion of the exact wave function  at small and large radial distances. Such expansion can be obtained by means of standard asymptotic analysis  or semiclassical considerations as it is explained in \cite{radialQAHO}. This simple and general prescription has been successfully used to construct wave functions of several quantum mechanical systems such as  anharmonic oscillators \cite{radialQAHO}, atomic \cite{TurbinerCompact} and molecular ones \cite{TURBINER2006309}, and complexes inside semiconductors \cite{delvalle2023two}. 

As a first step, it is convinient to obtain the Riccati equation from the radial Schr\"odinger equations (\ref{radialSeq}) by using the exponential substitution $R(r)=e^{-\Phi(r)}$. 
The Riccati equations read
\begin{equation}
    \label{Riccati}
\Phi^{\prime\prime}-\Phi^{\prime}\left(\Phi^\prime-\frac{2(1)}{r}\right)-2(v_0e^{-r^2}+E)=0\,,
\end{equation}
where coefficient (1) in parenthesis holds for 2D case. For large distances $r\to\infty$ the asymptotic behavior of $\Phi$ is 
\begin{equation}
    \label{phi_r->0}
    \Phi(r)=\sqrt{-2E}r+\varepsilon_1 \log(r)+ \frac{\varepsilon_3}{r} + O(1/r^2)\,,
\end{equation}
where $\varepsilon_1$ and $\varepsilon_3$ are coefficients.
For small distances $r\to0$ the asymptotic behavior of $\Phi$ is 
\begin{equation}
    \label{phi_r->0}
    \Phi(r)=\frac{1}{2}(v_0+E)r^2+\kappa_1 r^4 + O(r^6)\,
\end{equation}
where $\kappa_1$ is a coefficient.

Both asymptotic behaviors can be interpolated by the following simple Ansatz
\begin{equation}
\label{psitrial}
\phi(r,\chi)=\frac{(a + b r^2)}{\sqrt{1 + r^2}} + s \log \left( 1 + r^2\right)\,,
\end{equation}
where $\chi$ indicates a set of parameters $\chi=\{a,b,s\}$ which can be considered as variational parameters.

To compute the expectation value of the Hamiltonian and find the optimal variational parameters which minimize the variational energy we used a modification of  the FORTRAN code  described in \cite{TurbinerCompact}.

\section{Results}

\subsection{One particle in a Gaussian well}

We carried out numerical calculations to obtain the approximate eigenvalues and eigenstates of the Hamiltonian (\ref{HamOriginal}) by means of the LMM and variational method with the compact Anstaz (\ref{psitrial}). 
For reproducibility, the energy of the three lowest states is presented in Table I for different values of the effective parameter $v_0$. 

Far enough from the critical values, LMM provides twelve stable decimal figures confirmed by using the largest meshes considered in this work $N=1000$ and $N=2000$. In this regime a small mesh $N\sim50$ provides an accuracy of six decimal digits. 
On the other hand when $v_0$ approaches the critical value, the states become weakly bound and correspondingly the wave function becomes flat (see for example the expectation values $\langle r\rangle $ and its variance in Table \ref{tablaene}), 
such that the accuracy on the energy is dramatically deteriorated. Thus the access to the optimal wave functions close to the critical parameter is limited for computational resources. 

To address this problem, when the square normalizability of the wave function is compromised, a prevalent strategy involves maintaining the energy fixed at the threshold value (in this case, $E=0$) and subsequently solving the differential equation to determine the critical parameter. This methodology, as outlined in \cite{Liverts_2011} for instance, proves accurate in obtaining critical parameters for $s$-states. However, it is noteworthy that this approach does not provide information about the energy in close proximity to the threshold.

This study employs an alternative approach for estimating the critical parameter through the use of the LMM. To enhance accuracy as the critical threshold is approached, an extension of the configuration domain for the radial coordinate, denoted as $r$, is imperative. Within the framework of the LMM, this extension is achieved by augmenting the mesh size, represented by $N$.
The critical value, denoted as $v_0^c$, is approximated by iteratively decreasing $v_0$ in small increments until the energy of the selected state becomes positive. Various mesh sizes, denoted as $N$, were employed in this process. For states with angular momentum different from zero, the precision of the critical parameter $v_0^c$ obtained with $N=1000$ was corroborated with at least six decimal digits when employing $N=2000$. Table \ref{tablac} provides an overview of the critical parameters for the lowest states.

Concerning states with angular momentum equal to zero (s states), a discrepancy arises beyond the third decimal digit in the critical parameter $v_0^c$ when utilizing different meshes. Consequently, a more comprehensive analysis is warranted (refer to Appendix B for details). It is noteworthy that fixing the energy to zero yields higher accuracy for s states. However, for states with angular momentum $l>0$, the accuracy is sustained, and our approach exhibits improved precision as the angular momentum increases. This phenomenon is explicable by considering the form of the effective potential in the Schrödinger equation (\ref{radialSeq}), which differs for non-zero angular momentum.

Table \ref{tablap} presents the coefficients of the expansions (\ref{expansiona}) and (\ref{expansionb}) around the critical value for various states, obtained through interpolation and via the expectation value (\ref{HellmanFeynman}). Notably, for states with non-zero angular momentum, the coefficients of the leading term obtained through the expectation value coincide up to three decimal places when compared to those obtained through interpolation.

\begin{widetext}
 \begin{center}
\begin{table}[htb!]
\caption{\label{tablaene} Energies (in Hartrees), expectation values $\langle r \rangle $ (in units of Bohr radius) and variance $\sigma_r$ corresponding to the lowest states of the particle in the Gaussian well. Boldface indicates figures reproduced by the compact wave function (\ref{psitrial}).
}
\begin{center}
\begin{tabular}{cccccccccccc} \hline
\multicolumn{11}{c}{Three dimensions}  \\
\multicolumn{4}{c}{$(n=1,l=0)$} & \multicolumn{4}{c}{$(n=2,l=0)$} & \multicolumn{4}{c}{$(n=2,l=1)$} \\
$v_0$ & $E$ & $\langle r \rangle $ & $\sigma_r$ & $v_0$ & $E$ & $\langle r\rangle$& $\sigma_r$ & $v_0$ & $E$ & $\langle r \rangle $ & $\sigma_r$ \\ 
\hline
100  &  -{\bf 79.738\,800} & {\bf 0.314} & {\bf 0.135} & 100  & {\bf -55.388\,7}45  & {\bf 0.506} & {\bf 0.233}  &100 & -{\bf 66.896\,22}0 & {\bf 0.428} & {\bf 0.142} \\
50  &   -{\bf 35.958\,44}6 & {\bf 0.381} & {\bf 0.165} & 50   &   {\bf -19.987\,4}86 &  {\bf 0.641} & {\bf 0.292} &   50 & -{\bf 27.282\,42}8 & {\bf 0.526} & {\bf 0.178}  \\
10  &   -{\bf 4.280\,6}02  & {\bf 0.637} & {\bf 0.291} & 15   &  {\bf -1.214\,6}69 & {\bf 1.192}& {\bf 0.537}  & 10 & -{\bf 1.282\,8}61  & {\bf 0.997} & {\bf 0.398} \\
5 	&   -{\bf 1.271\,7}01  & {\bf 0.858} & {\bf 0.423} &12.5 &  {\bf -0.506}\,521 & {\bf 1.45}1 & {\bf 0.68}1 & 7.5& -{\bf0.361}\,014   & {\bf 1.25}8 & {\bf 0.5}88 \\ 
2	&   -{\bf 0.075}\,432  & {\bf 1.95}0 & {\bf 1.3}35 &10    &  {\bf -0.06}1\,198 & {\bf 2.5}25& {\bf 1}.520 & 6.5 &
-{\bf 0.08}9\,254 & {\bf 1.6}02 & {\bf 0} .945  \\
1.4	&   -{\bf 0.000}\,726  & {\bf 13}.834& {\bf 1}3.128 & 9     &  {\bf -0.0}00\,608 & {\bf 15}.527& {\bf 1}4.344 &6.1 &-{\bf 0.0}08\,031 & {\bf 2}.357 & {\bf 2}.164 \\ \hline
\multicolumn{11}{c}{Two dimensions}  \\
\multicolumn{4}{c}{$(n=1,l=0)$} & \multicolumn{4}{c}{$(n=2,l=0)$} & \multicolumn{4}{c}{$(n=2,l=1)$} \\ 
$v_0$ & $E$ & $\langle r \rangle $ & $\sigma_r$ & $v_0$ & $E$ & $\langle r\rangle$& $\sigma_r$ & $v_0$ & $E$ & $\langle r \rangle $ & $\sigma_r$ \\ \hline
100 & -{\bf 86.362\,354} &  {\bf 0.244} & {\bf 0.129}  & 100 & -{\bf 61.196\,987} & {\bf 0.456} & {\bf 0.222} & 100 & {\bf -73.249\,0}69 & {\bf 0.374} & {\bf 0.139} \\
10  & -{\bf 6.042 27}2   &   {\bf 0.476}& {\bf 0.26}1 & 10  & -{\bf 0.635\,5}48& {\bf 1.31}8 & {\bf 0.64}0 & 10  & {\bf -2.684}\,722 &  {\bf 0.80}1 & {\bf 0.33}0 \\
1   & -{\bf 0.115}\,386  & {\bf 1.5}81 & {\bf 1}.137  &  7  & -{\bf 0.04}1\,781 & {\bf 2.9}03 & {\bf 1}.873 & 5 &  {\bf-0.3}91\,199    &  {\bf 1.2}30 & {\bf 0}.618 \\
 \hline 
\end{tabular}
\end{center}
\end{table}
\end{center}
\end{widetext}

\begin{widetext}
 \begin{center}
\begin{table}[htb!]
\caption{\label{tablac} Critical parameters (in Hartrees) $v_0^c$ of the Gaussian wells in two and three dimensions for the lowest states ($n=1-4$ and $l=0-3$ corresponding to the standard notation s,p,d,f).
Digits market with boldface were obtained by extrapolation (\ref{fit}).  The second row marked with $\dagger$ stands for the results in \cite{Liverts_2011}.
}
\begin{center}
\begin{tabular}{c|cccccccc} \hline
\multicolumn{8}{c}{Three dimensions} \\
$n$/$l$ & $0$ & & $1$ & & $2$ & & $3$  \\ 
\hline
1  &  1.342\,{\bf 002} & &     -       & &     -            &  &     -   \\
   &  1.342  002\,3$^\dagger$      & &     -       & &     -            &  &     -   \\
2  &  8.897\,{\bf 850} & &  6.049\,655 & &     -            &  &     -   \\
   &  8.897\,     85$^\dagger$   & &  6.049\,654$^\dagger$  & &     -            &  &     -   \\
3  & 22.786\,{\bf 740} & & 17.544\,889 & & 13.450\,538\,800  &  &     -   \\
   & 22.786       74$^\dagger$   & & 17.544\,888$^\dagger$ & & 13.450\,539$^\dagger$  &  &     -   \\
4  & 42.981\,{\bf 700} & & 35.241\,429 & & 28.837\,886\,068 &  & 23.553\,939\,852 \\
   & 42.981\,     7$^\dagger$  & & 35.241\,428$^\dagger$ & & 28.837\,886$^\dagger$ &  & 23.553\,931$^\dagger$ \\
 \hline 
 \multicolumn{8}{c}{Two dimensions} \\
$n$/$l$ & $0$ & & $1$ & & $2$ & & $3$  \\ 
\hline
1  & - & & - & & -& & -\\
2  & 5.6{\bf 60} & & 3.359\,62  & &     -            &  &     -   \\
3  & 17.{\bf 71} & & 12.894\,53 & & 9.412\,85  &  &  -      \\

\end{tabular}
\end{center}
\end{table}
\end{center}
\end{widetext}

\begin{table}[htb!]
\caption{\label{tablap} Coefficients of the expansions (\ref{expansiona}), (\ref{expansionb}) and (\ref{C2d}) corresponding to the energy around the critical value. The coefficients marked with $\dagger$ correspond to those obtained via expectation values (\ref{HellmanFeynman}).
}
\begin{center}
\begin{tabular}{c|ccccc|c|ccccc} \hline
$(n,l)$ & $\gamma_2$ &  & $\gamma_3$  & & $\gamma_4$ & $(n,l)$ & $\xi_2$ & & $\xi_3$ & & $\xi_4$  \\ 
\hline
$(1,0)$ & -0.2212 & & 0.0966 & &-0.0682 & $(2,1)$ &-0.1422 & & -0.0734 & & -0.0081\\ 
 & & & & & & & -0.1423$^\dagger$ & & &\\ 
$(2,0)$ & -0.0593 & & 0.0101 & &-0.0002 & $(3,1)$ &-0.0794& & -0.0408& & -0.0048 \\
 & & & & & & & -0.0795$^\dagger$ & & &\\ 
$(3,0)$ & -0.0285 & & 0.0029 & &-0.0006 & $(3,2)$ &-0.0548 & &-0.0274 & &-0.0027\\
 & & & & & & & -0.0549$^\dagger$ & & &\\
$(4,0)$ & -0.0171 & & 0.0013 & &-0.0003 & $(4,1)$ &-0.2438 & &-0.0103 & &-0.0093 \\
 & \multicolumn{4}{c}{\textbf{Two dimensions}}& & & -0.2442$^\dagger$ & & &\\
 \cmidrule{2-4}
  &$\eta_1$ & & $\eta_2$ & & & $(4,2)$ &-0.1644 & & -0.0033 & & -0.0089\\
$(1,0)$& -4.4$\times$10$^{-4}$& & 0.2995 & & & & -0.1646$^\dagger$ &\\
$(2,0)$&  0.2712 & & & & & $(4,3)$&-0.2846 & & -0.0010& & -0.0058 \\
$(2,1)$& -0.2091 & & & & & & -0.2847$^\dagger$ &\\
 \hline 
\end{tabular}
\end{center}
\end{table}

\newpage

\subsection{Gaussian wells in nuclear physics}

Gaussian wells are recognized as valuable tool for characterizing the short-range interactions among nucleons in light nuclei \cite{CONTESSI2017839,Barnea2015}. The model that describe nuclei in the $s$-shell, where $A\leq4$, is outlined in \cite{Carleo2021} and the references therein
\begin{equation}
\label{hamnuc}
\hat{H}=-\hbar^2\sum_i\frac{\nabla^2_i}{2m_N}+\sum_{i<j}\left( C_1 + C_2 \vv{\sigma}_i \cdot \vv{ \sigma}_j \right) e^{-\Lambda^2 r_{ij}^2 /4}
+D_0\sum_{i<j<k}\sum_{cyc}e^{-\Lambda^2(r_{ik}+r_{ij})/4}\,,
\end{equation}
where $m_N$ is the nucleon mass and $\vv{\sigma}_i$ is the Pauli matrix acting on the $i$ nucleon spin. 
$\Lambda$, $D_0$, $C_1$ and $C_2$ are constants fitted to describe the Deuteron binding energy \cite{Pederiva2015}. Specifically, our attention is directed towards the Deuteron, which represents the simplest nucleus with $A=2$, comprising solely one neutron and one proton.
Since the potential in the Hamiltonian only depends on relative coordinates, the mass center coordinates can be separated. The spin variables of Deuteron can be integrated out by using spin eigenfunctions of $\vv{\sigma}_1\cdot\vv{\sigma}_2$ 
with eigenvalues $+1$ and $-3$.
Thus, the Hamiltonian in relative coordinates reads 
\begin{equation}
\label{Hdeuterio}
\hat{H}=-\frac{\hbar^2}{2\mu}\nabla_{r_{12}}^2 + (C_1+C_2)e^{-\Lambda^2 r_{12}^2/4}\,,
\end{equation}
where $\mu$ is the reduced mass and $r_{12}$ is the distance between the proton and neutron. The parameters of the Gaussian wells can be taken from the literature \cite{Pederiva2015} corresponding to $\Lambda=4\,{\rm fm}^{-1}$  ($C_1=-487.5\,$MeV, $C_2=-17.5\,$MeV) and  $\Lambda=6\,{\rm fm}^{-1}$ ($C_1=-1064\,$MeV, $C_2=-26\,$MeV). 
For adequate units of energy we use 
$$\frac{\hbar^2}{\mu}=\frac{\hbar^2c^2}{\mu c^2}=\frac{(197.3\, {\rm MeV}\cdot {\rm fm})^2}{469.5 \,{\rm MeV}}=82.9 \,{\rm MeV}\cdot {\rm fm}^2\,.$$

The calculation of the expectation value of the Hamiltonian (\ref{Hdeuterio}) is straight forward using the dimensionless trial wave function (\ref{psitrial}) with the substitution $r\to \Lambda r_{12}$ for adequate units of distance (fm). Additionally, variational calculations were conducted, employing a linear superposition of wave functions in the form (\ref{psitrial}), characterized by distinct sets of variational parameters \cite{Harris}. These parameters are designed to capture various configurations within the system
$$\Psi_{\chi}(r)=\sum_{i=1}^{\mathcal N} c_i e^{-\phi(r,{\chi_i})}\,$$
including up to three terms in the linear superposition (${\mathcal N}=1,2,3$)\footnote{In \cite{Harris}, it was found that only four configurations of a highly compact wave function yields to a competitive energy for the ground state of Helium.}.
The results for the ground state energy of deuteron are presented in Table \ref{tablanuclear} in comparison to the previous results obtained by Artifitial Neural Network correlator Ansatz \cite{Carleo2021} and conventional VMC calculations using a spline parametrization for the Jastrow functions \cite{Barnea2015}. 
A rapid convergence of the concise Ansatz (\ref{psitrial}) is evident, as only three terms (12 variational parameters) in the superposition yield competitive energies compared to those obtained through complex wave functions derived from sophisticated methods. Direct comparison with a Neural Networks wave function reveals that our approach requires only 12 parameters to achieve accuracy comparable to that achieved by a model with 24 parameters (6 hidden nodes) to reproduce 0.1 percent of the exact energy \cite{KEEBLE2020135743}. Notably, a Neural Networks wave function with 80 parameters (20 hidden nodes) is necessary to attain a level of accuracy equivalent to our 12-parameter wave function.

Given that the deuteron is situated within the weakly bound regime, where $\frac{mV_0}{\lambda\hbar^2}\sim \lambda_c$, the expansion (\ref{expansiona}) can be applied to estimate the binding energy of the deuteron for both values of $\Lambda$, i.e., $\Lambda=4,{\rm fm}^{-1}$ and $\Lambda=6,{\rm fm}^{-1}$. 
Identifying from Hamiltonian (\ref{Hdeuterio}), $V_{0} = C_{1} + C_{2}$ and $\lambda = \Lambda^{2}/4$, in such a way that the effective parameter of the potential takes the form $v_{0}= \frac{4 \mu (C_{1} + C_{2})}{\hbar^{2}\Lambda^{2}} $, the expression for Deuteron's ground state energy according to expansion (\ref{expansiona})
\begin{equation}
 \label{expansionDeutI}
 E_{1,0}= \frac{\Lambda^{2}\hbar^{2}}{4\mu}\sum_{n=2}^{\infty}\gamma_n\Big(\frac{4 \mu (C_{1} + C_{2})}{\hbar^{2}\Lambda^{2}} - v_{0}^{c}\Big)^{n}
\end{equation}
where coefficients $\gamma_{n}$ are taken from Table \ref{tablap}.

The corresponding results are presented in the last column of Table \ref{tablanuclear}. It is notable that the expansion (\ref{expansiona}) accurately describes two significant figures of the binding energy. Despite a slight overestimation of the energy, the associated error remains on the order of 0.3 percent concerning the most accurate results. Furthermore, the expansion (\ref{expansiona}) exhibits enhanced accuracy for $\Lambda=6,{\rm fm}^{-1}$ due to the closer proximity of the ratio to the critical value.

\begin{widetext}
 \begin{center}
\begin{table}[htb!]
\caption{\label{tablanuclear} Ground state energies in $MeV$ for the case of Deuteron. Dimensionless optimal variational parameters for the wave function are also included. The numerical  error on the variational energies are of the order $10^{-8}\,$MeV while the error for LMM energies with $N=2000$ is $\sim10^{-12}\,$MeV. 
}
{\small
\begin{center}
\begin{tabular}{c|ccccccc} \hline
& \multicolumn{6}{c}{Energy (MeV)}  \\
 $\Lambda$ (fm$^{-1}$) & 1 term & 2 terms & 3 terms  & LMM &  ANN-VMC  & VMC-JS & Eq. (\ref{expansionDeutI})  \\ 
\hline
$4$ & $-2.1639$ & $-2.2238$ & $-2.2242$ & $-2.2242$ &  $-2.224(1)$ & $-2.223(1)$ & $-2.2300$  \\
$6$ & $-2.1064$ & $-2.2232$ & $-2.2244$ & $-2.2244$ & $-2.224(4)$ & $-2.220(1)$ & $-2.2263$ \\ \hline
&\multicolumn{6}{c}{Optimal parameters }  \\
&\multicolumn{3}{c}{$\Lambda=4\,$fm$^{-1}$ } & \multicolumn{3}{c}{$\Lambda=6\,$fm$^{-1}$ }  \\ 
parameter & 1 term  & 2 terms & 3 terms & 1 term  & 2 terms & 3 terms  \\ \hline
 $a_1$ & $1.5031$& $0.9614$ & $0.9519$ & $1.3954$ & $0.8851$ & $0.8303$\\ 
 $b_1$ & $0.0349$& $0.0521$ & $0.0535$ & $0.0214$ & $0.0354$ & $0.0376$\\ 
 $s_1$ & $0.7284$& $0.5832$ & $0.5784$ & $0.7006$ & $0.5622$ & $0.5993$\\ 
 $a_2$ &         & $8.3627$ & $8.1823$ &          & $5.4551$ & $6.3339$\\ 
 $b_2$ &         & $0.8846$ & $0.9016$ &          & $1.1227$ & $1.2810$\\ 
 $s_2$ &         & $1.0675$ & $1.0127$ &          & $0.1445$ & $0.1832$\\ 
 $a_3$ &         &          & $60.0845$&          &          & $21.2916$ \\ 
 $b_3$ &         &          & $0.1164$ &          &          & $0.0141$ \\ 
 $s_3$ &         &          & $0.8402$ &          &          & $1.4484$ \\ 
\end{tabular}
\end{center}
}
\end{table}
\end{center}
\end{widetext}

 \clearpage

 \subsection{The electronic structure of quantum dots}
 
\begin{figure}
    \centering
    \begin{tikzpicture}
    \coordinate (A) at (0,0);
    \coordinate (B) at (0,3);
    \coordinate (C) at (3*0.866,-3*0.5); 

    \fill (B) circle (1.5pt);
    \fill (C) circle (1.5pt);
    \node[above] at (B) {$e_2$};
    \node[right, below] at (C) {$e_1$};

    \draw[-,thick] (A) -- node[left] {$\mathbf{r_2}$} (B);
    \draw[-,thick] (A) -- node[below] {$\mathbf{r_1}$} (C);
    \draw[-,dashed](C)--node[right]{$\mathbf{r}_{12}$}(B);
    
    \pic[draw,"$\theta_1$",  angle radius=3.5mm, pic text options={above=5pt, right=5pt}]{angle=C--A--B};
    \end{tikzpicture}
    \caption{Illustration of a two-electron quantum dot. The reference frame is positioned such that $\mathbf{r}_{2}$ aligns with the z-axis. This simplifies the consideration to just the angle $\theta_{1}$ between both vectors, rather than the pair of angles associated with each vector.}
    \label{fig: quantum dot}
\end{figure}
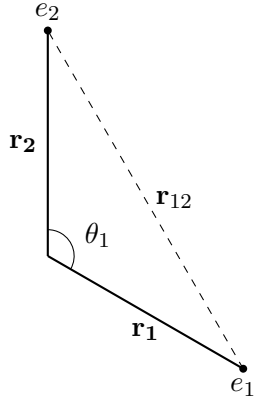

Lets consider the case of two electrons in a quantum dot where the confinement potential is modeled by a Gaussian well \cite{QUANTUMDOTS}
\begin{equation}\label{eq: 2D quantum dot}
\hat{H}=-\frac{\hbar}{2m}\left(\nabla_1^2+\nabla_2^2\right)-V_0e^{-\lambda r_1^2}-V_0e^{-\lambda r_2^2}+\frac{e^2}{r_{12}},
\end{equation}
where $\nabla^2_i \quad i=1,2$ is the three dimensional Laplacian of the i-th electron.
Beyond the one-body potential in form of Gaussian wells, there is an inter-electronic Coulombian repulsion $1/r_{12}$. Therefore the wave function is subject to satisfy the Cusp condition 
\cite{Kato1951FundamentalPO} at the electron-electron coalescence $\langle \delta({\bf r_{12}})\partial_{r_{12}}\rangle /\langle \delta({\bf r_{12}})\rangle\sim1/2$.
Analogous to the classical two-body problem, the physics of the two-electron quantum dot is constrained to a plane. Hence, although the vector associated to each particle exists in three dimensions, the degree of freedom of the problem remains three under an appropriate reference frame. This can be achieved by selecting a reference frame such that $\theta_2=\pi/2$ as shown in Figure \ref{fig: quantum dot}, reducing the wave function effective dependence to three coordinates: $r_1$, $r_2$, and $\theta_1$, being the later the angle between the two vectors. The Hamiltonian \eqref{eq: 2D quantum dot} does not admit an analytical solution, and similar to the one body case, it possesses a finite number of bound states which have been explored extensively in \cite{QUANTUMDOTS}. 
Under the reescaling $r_i\to\lambda^{1/2}r_i$, the Hamiltonian \eqref{eq: 2D quantum dot} transforms as
\begin{equation}
\label{Htransformado}
\frac{1}{\lambda}\hat{H}=-\frac{\hbar}{2m}\left(\nabla_1^2+\nabla_2^2\right)-\frac{V_0}{\lambda}(e^{ -r_1^2}+e^{-r_2^2})+\frac{e^2}{\sqrt{\lambda}r_{12}}\,.
\end{equation}
This transformation is the analogous to that which generates the $1/Z$ expansion of Helium \cite{Zexpansion}. 
Now we can use our compact wave function (\ref{psitrial}) as orbitals following the one-body terms in (\ref{Htransformado}). The wave function for the $1s^2$ state of the two-electron quantum dot can be build as the product of one orbital (\ref{psitrial}) for each electron and a Jastrow factor depending on the relative distance
\begin{equation}
\label{trialQD}
    \psi(r_1,r_2,\theta_1,\chi_2)=\exp\left(-\phi(\alpha^2\,r_1,\chi_0)-\phi(\beta^2\,r_2,\chi_0)\right)(1+cr_{12})\exp\left(\gamma\,r_{12}*\frac{1+\delta_{1}r_{12}}{1+\delta_2r_{12}}\right)\,,
\end{equation}
where $\chi_0$ are the variational parameters for the one particle case, for a given effective parameter $v_0=V_0/\lambda$. Here we introduce a new set of variational parameters, namely, $\chi_2=\{\alpha,\beta,c,\gamma,\delta_1,\delta_2\}$. The parameters $\alpha$ and $\beta$ will play a role of a screening parameter, while the rest of parameters describe the electron-electron interaction \cite{TurbinerCompact}.
Furthermore we can impose the constraint \cite{TurbinerCompact} $c=1/2-\gamma$ in order to guarantee that the wave function fulfill the cusp condition. Thus the total number of variational parameter is five. The results are displayed and Table \ref{resultsQDot} and they show that our few-parametric wave function (\ref{trialQD}) yields to competitive results against those reported in the literature while the cusp condition is fullfilled, simultaneously. Occasionally even lower energies are obtained. To understand this behavior, we include the expected value of the inverse of the inter-electronic interaction term $1/r_{12}$. As shown in Table \ref{tab: r12}, our variational results outperform the numerical results \cite{QUANTUMDOTS} when the inter-electronic repulsion is small while the accuracy deteriorates for large inter electronic distances.  

\begin{widetext}
\begin{center}
\begin{table}[htb!]
\caption{\label{resultsQDot} Total energy and expectation values of the ground state state $1s^{2 \,1}S$ of the two electron Gaussian Quantum Dot. Superindex $\dagger$ indicates results in \cite{QUANTUMDOTS}. Bold numbers indicates the lowest energy. Al quantities are presented in atomic units. }
\begin{center}
\begin{tabular}{cc|cc|c} \hline
 $\lambda$ & $V_0$ & $E_T$  & $E_T^\dagger$  &$\Big<\frac{1}{r_{12}}\Big>$\\ \hline
0.05  & 10 & -16.380161 & \textbf{-16.389353} & 0.152650 \\
      &  8 & -12.740711 & \textbf{-12.750555} & 0.143053\\
      &  6 & -9.155680 & \textbf{-9.164120} & 0.131107\\ \hline
0.1   & 10 & \textbf{-15.110462} & -15.110283 & 0.252596\\ 
      &  8 & -11.611217 & \textbf{-11.612391} & 0.237427\\
      &  6 & -8.186968 & \textbf{-8.186970} & 0.215436\\ \hline
0.2   & 10 & \textbf{-13.406999} & -13.397977 & 0.415227 \\
      &  8 & \textbf{-10.108105} & -10.098433 & 0.384509 \\
      &  6 & \textbf{-6.909523} & -6.899916 & 0.348081\\ \hline
0.5   & 10 & \textbf{-10.3079923} & -10.290023 & 0.781601 \\
      &  8 & \textbf{-7.402543}& -7.385087 &0.719914   \\
\end{tabular}
\end{center}
\label{tab: r12}
\end{table}
\end{center}
\end{widetext}

\section{CONCLUSIONS}

The problem of a particle confined in a two and three-dimensional Gaussian well is addressed through two distinct methods: i) a highly accurate numerical approach (LMM) and ii) a semi-qualitative but locally accurate compact Ansatz.

The precision afforded by the LMM enables a thorough exploration of the spectrum, spanning from the strongly bound regime to the weakly bound regime, facilitating the estimation of critical parameters. The number of bound states within a Gaussian well is determined by the parameter ratio $v_0=V_0/\lambda$. In the case of three-dimensional Gaussian wells, those with $v_0$ values below the critical threshold, approximately $v_0^c\sim1.342$, do not contain bound states.

The introduced compact Ansatz in this study exhibits local accuracy and accurately describes at least three significant digits across the entire parameter domain, utilizing merely three variational parameters. The convergence rate of this Ansatz is scrutinized within a realistic nuclear physics model for deuterium, incorporating Gaussian interaction to portray short-range nucleon interactions and spin terms in the Hamiltonian. The variational energy displays an exceedingly rapid convergence rate when employing a superposition of wave functions, sharing the same form but differing in parameter configurations. Moreover, an analytical expression for the binding energy of the Deuteron was formulated through the expansion of the energy in the vicinity of the critical parameter. The derived formula exhibits notably high accuracy, particularly as the effective parameter approaches critical values. This formula open the door for a simpler look at the weakly binding regime where the accuracy of all approximation methods dramatically deteriorates. 

The rapid convergence of the wave function can be exploited in the exploration of the electronic structure of quantum dots. In a quick glance to the problem of with a two-electron quantum dot with Gaussian one-body potential, the few-parametric wave function provides competitive results against the most accurate results in the literature. This underscores the efficacy of constructing a few-parametric wave function guided by physical motivation in contrast to extensive expansions within a complete basis.

\section{Acknowledgments}

DJN thanks the kind hospitality of the Brown University Department of Chemistry. This project was partially supported by Fulbright COMEXUS Project NO. P000003405. 
The authors also thank J.C. del Valle for his help with the LMM and useful discussions.


\section*{Appendix A: The Lagrange Mesh Method}\label{Appendix A}
The Lagrange Mesh Method (LMM)  belongs to the family of pseudospectral and collocation methods using a non uniform mesh, which is defined by the zeroes of a given orthogonal polynomial. In particular, in the formulation presented in \cite{BAYEPhysRep}, the LMM is an accurate numerical method exploiting the Gauss quadrature to evaluate matrix elements and expectation values of any operator. It is worth mentioning that pseudospectral methods have shown to be adequate for computing with extremely high accuracy critical behavior while other methods fail, see e.g. \cite{Jiao_2022}. As a first step in the LMM, the radial part of the wave function $R_{n,l}(r)$ is assumed as a linear superposition of the regularized Lagrange orthonormal\footnote{Under the Gauss quadrature $\hat{f}_i(r)$ are orthogonal.} functions $\{\hat{f}_i(r)\}_{i=1}^N$,
\begin{equation}
\label{psi}
R_{n,l}(r)=r^{-1}\sum_{i=1}^N c_i\,\hat{f}_i(r)\ , \quad \sum_{i=1}^N|c_i|^2=1\ ,
\end{equation} 
where $c_i$, $i=1,2,...N$, are real coefficients. The polynomials to build the Lagrange functions are chosen following the 
form of the kinetic energy and the domain of the coordinate \cite{BAYEPhysRep}. We describe the convenient Lagrange functions for each case in the following subsections. In any case, the secular Schr\"odinger equations reads
\begin{equation}
\label{secular}
\sum_{j=1}^{N} \left\{T_{ij}\ +\ U_{ij}\ -\ \delta_{ij}E\right\}c_i\ =\ 0\ ,\end{equation}
and its solution determines the approximate energies and the coefficients $c_i$'s of the wave function. The matrix elements of the effective potential can be easily obtained by means of the Gauss quadrature
\begin{eqnarray}
\label{matrixV}
U_{ij}\ &=& \ \left(-v_0 e^{-r_i^2}+\ \frac{(d+2 \ell-1) (d+2 \ell-3)}{8r^2} \right)\delta_{ij}\,.\, \nonumber 
\end{eqnarray} 
However, the matrix elements corresponding to the kinetic energy $T_{ij}\,$ depend on the choice of the mesh since the operator contain derivatives and will be discussed in the next subsection as well. 

 \subsection{Two Dimensional Case}

 For two dimensions, notice that the form of the effective potential in (\ref{radialSeq}) changes drastically with the angular momentum
 $l$.  For the ground state with angular momentum $l=0$, the divergence on the effective potential is negative, and therefore it is necesary a regularization to circumvent the problem of the negative singularity. In this case, we can consider the kinetic energy in the form
 $$\hat{T}=-\frac{d^2}{dr^2}+\frac{\alpha(\alpha-2)}{4r^2}\,,$$
 with $\alpha=1$. In this case, the convenient choice for building the mesh are the generalized Laguerre polynomials $L^\alpha_N$ where $N$ is the degree of the Laguerre polynomial $L^\alpha_N(r)$ and therefore the dimension of the non-uniform mesh.  The set $\{r_i\}_{i=1}^N$ corresponds to the $N$ real roots of the polynomial $L^\alpha_N(r)$.  The generalized Lagrange functions are given by
\begin{equation}
\label{lagrangefunctionaslpha}
\hat{f}_i(r)=(-1)^i r_i^{1/2}(\nu^\alpha_N)^{-1/2} \frac{L_N^\alpha (r)}{r-r_i} r^{\alpha/2} e^{-r/2}\,, 
\end{equation}
where $$\nu_N^\alpha=\frac{\Gamma(N+\alpha+1)}{N!}\,,$$ and the matrix elements of the kinetic energy operator are the following \cite{BAYEPhysRep}
\begin{eqnarray}
    T_{ii}&=& -\frac{1}{12} + \frac{(2N+\alpha+1)(\alpha+4)}{6(\alpha+1)r_i} +\frac{(\alpha+2)(\alpha-5)}{6r_i^2} ,\quad\quad i= j\ ,\\
    T_{ij} &=& (-1)^{i-j}\frac{1}{\sqrt{r_ir_j}}\left[ \frac{N}{\alpha+1} + \frac{1}{2} - \left(  \frac{1}{r_i} + \frac{1}{r_j} \right) + \frac{r_i+r_j}{(r_i-r_j)^2} \right] \ ,\quad\quad i\neq j\ .
\end{eqnarray}

For $l>0$, where the divergence in $\sim1/r^2$ is positive, we can consider the standard form of the kinetic energy 
 $\hat{T}=-\frac{d^2}{dr^2}$ i.e. the Laguerre mesh is more adequate since the centrifugal term can be considered as part of the potential. The Lagrange mesh is be discussed in the next subsection for the three-dimensional case.

\subsection{Three-dimensional Case}

For the three dimensional case, it is convenient to chose the Laguerre mesh given by
\begin{equation}
\hat{f}_i(r)=\ \frac{(-1)^i}{r_i^{1/2}}\,\frac{r\,L_N(r)}{(r-r_i)} \,e^{-r/2}\,. 
 \end{equation}
 For the particular case of the Laguerre mesh, the matrix elements of the kinetic term are \cite{BAYEPhysRep}
\begin{eqnarray}
T_{ii}\ &=&\ \frac{4\ +\ (4N\ +\ 2)\,r_i\ -\ r_i^2}{24\,r_i^2}\ ,\quad\quad i= j\ ,\\
T_{ij}\ &=&\ \frac{(-1)^{i-j}\,(r_i\ +\ r_j)}{2\,(r_i\, r_j)^{1/2}\,(r_i \ -\ r_j)^2}\ ,\quad\quad i\neq j\ .
\end{eqnarray}

Once the matrix representation is built, the secular equation is solved via diagonalization. The approximate energies are the eigenvalues and the eigenvectors give us the coefficients $c_i$ of the wave function (\ref{psi}). 
Then the expectation values of any local operator $g(r)$ in the state $(n,l)$ can be approximated by
 \begin{equation}
 \label{transition}
  \langle \psi_{n,l} |g(r)|\psi_{n,l}\rangle
 \ \approx\ \sum_{i=1}^{N} |c_i^{(n,l)}|^2\,g(r_i)\,.
 \end{equation} 
 
 In practice the numerical calculations can be performed with the Mathematica package \cite{delValleLMM}.

 The Lagrange Mesh Method (LMM) is a very efficient numerical method to solve the Schr\"odinger equation \cite{BAYEPhysRep,Baye_2011}, via the secular equation 
\begin{equation}
\label{secular}
\sum_{j=1}^{N} \left\{T_{ij}\ +\ U_{ij}\ -\ \delta_{ij}E\right\}c_i\ =\ 0,
\end{equation}
where $T_{ij}$ and $U_{ij}$ are the matrix elements of the kinetic and potential terms respectively in the Lagrange basis (for a more detailed discussion of the method, see \cite{BAYEPhysRep}). In this work, we make very similar considerations as in \cite{delvalle2023two}, being the matrix elements of the effective potential 
\begin{eqnarray}
\label{matrixV}
U_{ij}\ &=& \ \left(-v_0 e^{-r_i^2}+\ \frac{(d+2 \ell-1) (d+2 \ell-3)}{8r^2} \right)\delta_{ij}\,.\, \nonumber 
\end{eqnarray} 
In turn, for the kinetic elements we have to make a distinction between the two dimensional case and the third dimensional one. For the ground state in two dimensions, the matrix elements $T_{ij}$ are \cite{BAYEPhysRep}
\begin{eqnarray}
    T_{ii}&=& -\frac{1}{12} + \frac{(2N+\alpha+1)(\alpha+4)}{6(\alpha+1)r_i} +\frac{(\alpha+2)(\alpha-5)}{6r_i^2} ,\quad\quad i= j\ ,\\
    T_{ij} &=& (-1)^{i-j}\frac{1}{\sqrt{r_ir_j}}\left[ \frac{N}{\alpha+1} + \frac{1}{2} - \left(  \frac{1}{r_i} + \frac{1}{r_j} \right) + \frac{r_i+r_j}{(r_i-r_j)^2} \right] \ ,\quad\quad i\neq j\,
\end{eqnarray}
due to the negative divergence on the effective potential for $l=0$ on (\eqref{Schro2D}). For the rest of the states we can consider the standard form of the kinetic energy $\hat{T}=-\frac{d^2}{dr^2}$. For the three dimensional case, it is convenient to chose the Laguerre mesh being the matrix elements \cite{BAYEPhysRep}
\begin{eqnarray}
T_{ii}\ &=&\ \frac{4\ +\ (4N\ +\ 2)\,r_i\ -\ r_i^2}{24\,r_i^2}\ ,\quad\quad i= j\ ,\\
T_{ij}\ &=&\ \frac{(-1)^{i-j}\,(r_i\ +\ r_j)}{2\,(r_i\, r_j)^{1/2}\,(r_i \ -\ r_j)^2}\ ,\quad\quad i\neq j\ .
\end{eqnarray}
Then the expectation values of any local operator $g(r)$ in the state $(n,l)$ can be approximated by
 \begin{equation}
 \label{transition}
  \langle \psi_{n,l} |g(r)|\psi_{n,l}\rangle
 \ \approx\ \sum_{i=1}^{N} |c_i^{(n,l)}|^2\,g(r_i)\,.
 \end{equation} 
 
 In practice the numerical calculations can be performed with the Mathematica package \cite{delValleLMM}.

\section*{Appendix B: States with zero angular momentum}\label{appendix B}
It is well known that the wave functions  of states with zero angular momentum are particularly extended in the radial direction \cite{delValle2018} close to the threshold $E\to 0$, see for instance the variance of the radial coordinate in Table \ref{tablaene}. Physically, this is a consequence of the absence of centrifugal term in the effective potential. The mesh with $N=2000$ points is not large enough to confirm four decimal figures on the critical values of $v_0$. Therefore we have to introduce a scaling parameter $r_i\to hr_i$ which help to fit the point mesh distribution \cite{BAYEPhysRep} according to the physical problem. In particular if $h>1$ the mesh points are distributed along a larger domain. 
In principle the energy should be independent of $h$, however it can be treated as variational parameter \cite{BAYEPhysRep}.
In practice one can notice that the optimal value of $h$ increase as the parameter approaches the critical value. As a consequence, the energy decrease as a function of $h$ and therefore the critical parameter is shifted as well. It can be expected that, for large values of $h$, the mesh should reproduce correctly the behavior of the flat wave function and the estimate of the critical value $v_0^c$ should be more accurate.
In Figure 1, we plot the value of the parameter $v_0^c$ for which the energy of the ground state become positive as a function of the scaling parameter $h$, for different meshes $N=500,700,1000$. It can be seen that the three curves go asymptotically to the same value for $h\to\infty$.
Needless to point out that the larger the mesh the fastest the curve reaches its asymptotic value. However if $h$ is too large, the mesh points are too far from each other and the approximation is suddenly broken i.e. we have no access to the approximate energy for large values $h$.
Nevertheless, for each curve we can interpolate the the numerical results for $v_0^c$, using a fitting function in the form of a series in negative powers of $h$
\begin{equation}
 \label{fit}
 f(h)=\sum_{n=0}^3 \beta_n h^{-n\tau}\,,
\end{equation}
 where $\beta_n$ and $\tau$ are fitting parameters. The fit confirms that $\tau$ is close to the unity. Then, the estimate of the critical parameter is given as the leading term in (\ref{fit}) i.e. $v_0^c(h\to\infty)\approx \beta_0$.
 We consider only stable decimal digits of $\beta_0$ which are confirmed  with the largest meshes $N=1000$ and $N=2000$.
In Table \ref{tablac}, we present three additional figures in boldface for the critical parameter of each state, obtained by using the fit (\ref{fit}).

\begin{figure}[!h]
\begin{center}
\includegraphics[angle=0,width=100mm]{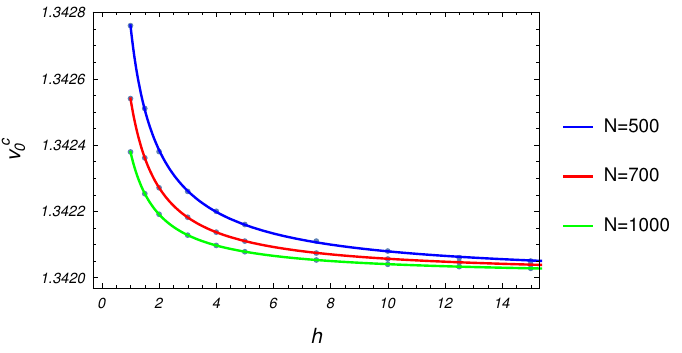}
\caption{\label{energyfit} Critical parameters  $v_0^c$ for the ground state in three dimensions as a function of the scaling parameter $h$. Each curve represents the critical value obtained by using different sizes of the mesh $N=500,700,1000$. Points correspond to the numerical results and solid curves correspond to the fit (\ref{fit}).
}
\end{center}
\end{figure}

\bibliography{Qdots2020}

\begin{thebibliography}{49}
\expandafter\ifx\csname natexlab\endcsname\relax\def\natexlab#1{#1}\fi
\expandafter\ifx\csname bibnamefont\endcsname\relax
  \def\bibnamefont#1{#1}\fi
\expandafter\ifx\csname bibfnamefont\endcsname\relax
  \def\bibfnamefont#1{#1}\fi
\expandafter\ifx\csname citenamefont\endcsname\relax
  \def\citenamefont#1{#1}\fi
\expandafter\ifx\csname url\endcsname\relax
  \def\url#1{\texttt{#1}}\fi
\expandafter\ifx\csname urlprefix\endcsname\relax\def\urlprefix{URL }\fi
\providecommand{\bibinfo}[2]{#2}
\providecommand{\eprint}[2][]{\url{#2}}

\bibitem[{\citenamefont{Fernandez}(2011)}]{FernandezGaussianWell}
\bibinfo{author}{\bibfnamefont{F.~M.} \bibnamefont{Fernandez}},
  \bibinfo{journal}{Amer. J. Phys.} \textbf{\bibinfo{volume}{79}},
  \bibinfo{pages}{752} (\bibinfo{year}{2011}).

\bibitem[{\citenamefont{Nandi}(2010)}]{NandiGaussianWells}
\bibinfo{author}{\bibfnamefont{S.}~\bibnamefont{Nandi}},
  \bibinfo{journal}{Amer. J. Phys.} \textbf{\bibinfo{volume}{78}},
  \bibinfo{pages}{1341} (\bibinfo{year}{2010}).

\bibitem[{\citenamefont{Descouvemont and Baye}(2010)}]{Descouvemont_2010}
\bibinfo{author}{\bibfnamefont{P.}~\bibnamefont{Descouvemont}}
  \bibnamefont{and} \bibinfo{author}{\bibfnamefont{D.}~\bibnamefont{Baye}},
  \bibinfo{journal}{Rep. Prog. Phys.} \textbf{\bibinfo{volume}{73}},
  \bibinfo{pages}{036301} (\bibinfo{year}{2010}).

\bibitem[{\citenamefont{Xiao and Xiao}(2019)}]{Xiao_2019}
\bibinfo{author}{\bibfnamefont{W.}~\bibnamefont{Xiao}} \bibnamefont{and}
  \bibinfo{author}{\bibfnamefont{J.}~\bibnamefont{Xiao}}, \bibinfo{journal}{J.
  Semicond} \textbf{\bibinfo{volume}{40}}, \bibinfo{pages}{042901}
  (\bibinfo{year}{2019}).

\bibitem[{\citenamefont{Gylfadottir et~al.}(2006)\citenamefont{Gylfadottir,
  Harju, Jouttenus, and Webb}}]{Gylfadottir2006}
\bibinfo{author}{\bibfnamefont{S.~S.} \bibnamefont{Gylfadottir}},
  \bibinfo{author}{\bibfnamefont{A.}~\bibnamefont{Harju}},
  \bibinfo{author}{\bibfnamefont{T.}~\bibnamefont{Jouttenus}},
  \bibnamefont{and} \bibinfo{author}{\bibfnamefont{C.}~\bibnamefont{Webb}},
  \bibinfo{journal}{New J. Phys.} \textbf{\bibinfo{volume}{8}},
  \bibinfo{pages}{211} (\bibinfo{year}{2006}).

\bibitem[{\citenamefont{Cuevas et~al.}(2013)\citenamefont{Cuevas, Kevrekidis,
  Malomed, Dyke, and Hulet}}]{Cuevas_2013}
\bibinfo{author}{\bibfnamefont{J.}~\bibnamefont{Cuevas}},
  \bibinfo{author}{\bibfnamefont{P.~G.} \bibnamefont{Kevrekidis}},
  \bibinfo{author}{\bibfnamefont{B.~A.} \bibnamefont{Malomed}},
  \bibinfo{author}{\bibfnamefont{P.}~\bibnamefont{Dyke}}, \bibnamefont{and}
  \bibinfo{author}{\bibfnamefont{R.~G.} \bibnamefont{Hulet}},
  \bibinfo{journal}{New J. Phys.} \textbf{\bibinfo{volume}{15}},
  \bibinfo{pages}{063006} (\bibinfo{year}{2013}).

\bibitem[{\citenamefont{Chaudhuri}(2021)}]{Chaudhuri}
\bibinfo{author}{\bibfnamefont{S.}~\bibnamefont{Chaudhuri}},
  \bibinfo{journal}{Physica E Low Dimens. Syst. Nanostruct.}
  \textbf{\bibinfo{volume}{128}}, \bibinfo{pages}{114571}
  (\bibinfo{year}{2021}).

\bibitem[{\citenamefont{Adamowski et~al.}(2000)\citenamefont{Adamowski,
  Sobkowicz, Szafran, and Bednarek}}]{PhysRevB.62.4234}
\bibinfo{author}{\bibfnamefont{J.}~\bibnamefont{Adamowski}},
  \bibinfo{author}{\bibfnamefont{M.}~\bibnamefont{Sobkowicz}},
  \bibinfo{author}{\bibfnamefont{B.}~\bibnamefont{Szafran}}, \bibnamefont{and}
  \bibinfo{author}{\bibfnamefont{S.}~\bibnamefont{Bednarek}},
  \bibinfo{journal}{Phys. Rev. B} \textbf{\bibinfo{volume}{62}},
  \bibinfo{pages}{4234} (\bibinfo{year}{2000}).

\bibitem[{\citenamefont{Boyacioglu et~al.}(2007)\citenamefont{Boyacioglu,
  Saglam, and Chatterjee}}]{Boyacioglu_2007}
\bibinfo{author}{\bibfnamefont{B.}~\bibnamefont{Boyacioglu}},
  \bibinfo{author}{\bibfnamefont{M.}~\bibnamefont{Saglam}}, \bibnamefont{and}
  \bibinfo{author}{\bibfnamefont{A.}~\bibnamefont{Chatterjee}},
  \bibinfo{journal}{J. Condens. Matter Phys.} \textbf{\bibinfo{volume}{19}},
  \bibinfo{pages}{456217} (\bibinfo{year}{2007}).

\bibitem[{\citenamefont{Gomez and Romero}(01 Mar.
  2009)}]{FewelectronsemiconductorquantumdotswithGaussianconfinement}
\bibinfo{author}{\bibfnamefont{S.}~\bibnamefont{Gomez}} \bibnamefont{and}
  \bibinfo{author}{\bibfnamefont{R.}~\bibnamefont{Romero}},
  \bibinfo{journal}{Open Physics} \textbf{\bibinfo{volume}{7}},
  \bibinfo{pages}{12 } (\bibinfo{year}{01 Mar. 2009}).

\bibitem[{\citenamefont{Xie}(2008)}]{XIE20082828}
\bibinfo{author}{\bibfnamefont{W.}~\bibnamefont{Xie}},
  \bibinfo{journal}{Physica B Condens. Matter} \textbf{\bibinfo{volume}{403}},
  \bibinfo{pages}{2828 } (\bibinfo{year}{2008}).

\bibitem[{\citenamefont{Poszwa}(2016)}]{Poszwa2016}
\bibinfo{author}{\bibfnamefont{A.}~\bibnamefont{Poszwa}},
  \bibinfo{journal}{Few-Body. Syst.} \textbf{\bibinfo{volume}{57}},
  \bibinfo{pages}{1127} (\bibinfo{year}{2016}).

\bibitem[{\citenamefont{Nader et~al.}(2017)\citenamefont{Nader,
  Alvarez-Jiménez, and Mejía-Díaz}}]{Nader2017}
\bibinfo{author}{\bibfnamefont{D.~J.} \bibnamefont{Nader}},
  \bibinfo{author}{\bibfnamefont{J.}~\bibnamefont{Alvarez-Jiménez}},
  \bibnamefont{and}
  \bibinfo{author}{\bibfnamefont{H.}~\bibnamefont{Mejía-Díaz}},
  \bibinfo{journal}{Few-Body. Syst.} \textbf{\bibinfo{volume}{58}},
  \bibinfo{pages}{1} (\bibinfo{year}{2017}).

\bibitem[{\citenamefont{Barnea et~al.}(2015)\citenamefont{Barnea, Contessi,
  Gazit, Pederiva, and van Kolck}}]{Barnea2015}
\bibinfo{author}{\bibfnamefont{N.}~\bibnamefont{Barnea}},
  \bibinfo{author}{\bibfnamefont{L.}~\bibnamefont{Contessi}},
  \bibinfo{author}{\bibfnamefont{D.}~\bibnamefont{Gazit}},
  \bibinfo{author}{\bibfnamefont{F.}~\bibnamefont{Pederiva}}, \bibnamefont{and}
  \bibinfo{author}{\bibfnamefont{U.}~\bibnamefont{van Kolck}},
  \bibinfo{journal}{Phys. Rev. Lett.} \textbf{\bibinfo{volume}{114}},
  \bibinfo{pages}{052501} (\bibinfo{year}{2015}).

\bibitem[{\citenamefont{Roy}(2016)}]{https://doi.org/10.1002/qua.25108}
\bibinfo{author}{\bibfnamefont{A.~K.} \bibnamefont{Roy}},
  \bibinfo{journal}{Int. J. Quantum Chem.} \textbf{\bibinfo{volume}{116}},
  \bibinfo{pages}{953} (\bibinfo{year}{2016}).

\bibitem[{\citenamefont{del Valle and Nader}(2018)}]{delValle2018}
\bibinfo{author}{\bibfnamefont{J.~C.} \bibnamefont{del Valle}}
  \bibnamefont{and} \bibinfo{author}{\bibfnamefont{D.~J.} \bibnamefont{Nader}},
  \bibinfo{journal}{J. Math. Phys.} \textbf{\bibinfo{volume}{59}},
  \bibinfo{pages}{102103} (\bibinfo{year}{2018}).

\bibitem[{\citenamefont{Gomes et~al.}(1994)\citenamefont{Gomes, Chacham, and
  Mohallem}}]{PhysRevA.50.228}
\bibinfo{author}{\bibfnamefont{O.~A.} \bibnamefont{Gomes}},
  \bibinfo{author}{\bibfnamefont{H.}~\bibnamefont{Chacham}}, \bibnamefont{and}
  \bibinfo{author}{\bibfnamefont{J.~R.} \bibnamefont{Mohallem}},
  \bibinfo{journal}{Phys. Rev. A} \textbf{\bibinfo{volume}{50}},
  \bibinfo{pages}{228} (\bibinfo{year}{1994}).

\bibitem[{\citenamefont{Edwards et~al.}(2017)\citenamefont{Edwards, Gerber,
  Schubert, Trejo, and Weber}}]{10.1093/ptep/ptx107}
\bibinfo{author}{\bibfnamefont{J.~P.} \bibnamefont{Edwards}},
  \bibinfo{author}{\bibfnamefont{U.}~\bibnamefont{Gerber}},
  \bibinfo{author}{\bibfnamefont{C.}~\bibnamefont{Schubert}},
  \bibinfo{author}{\bibfnamefont{M.~A.} \bibnamefont{Trejo}}, \bibnamefont{and}
  \bibinfo{author}{\bibfnamefont{A.}~\bibnamefont{Weber}},
  \bibinfo{journal}{Prog. Theor. Exp. Phys.} \textbf{\bibinfo{volume}{2017}}
  (\bibinfo{year}{2017}), \bibinfo{note}{083A01}.

\bibitem[{\citenamefont{Stubbins}(1993)}]{PhysRevA.48.220}
\bibinfo{author}{\bibfnamefont{C.}~\bibnamefont{Stubbins}},
  \bibinfo{journal}{Phys. Rev. A} \textbf{\bibinfo{volume}{48}},
  \bibinfo{pages}{220} (\bibinfo{year}{1993}).

\bibitem[{\citenamefont{Diaz et~al.}(1991)\citenamefont{Diaz, Fernandez, and
  Castro}}]{Diaz_1991}
\bibinfo{author}{\bibfnamefont{C.~G.} \bibnamefont{Diaz}},
  \bibinfo{author}{\bibfnamefont{F.~M.} \bibnamefont{Fernandez}},
  \bibnamefont{and} \bibinfo{author}{\bibfnamefont{E.~A.}
  \bibnamefont{Castro}}, \bibinfo{journal}{J. Phys. A}
  \textbf{\bibinfo{volume}{24}}, \bibinfo{pages}{2061} (\bibinfo{year}{1991}).

\bibitem[{\citenamefont{Serra et~al.}(1998)\citenamefont{Serra, Neirotti, and
  Kais}}]{doi:10.1021/jp9820572}
\bibinfo{author}{\bibfnamefont{P.}~\bibnamefont{Serra}},
  \bibinfo{author}{\bibfnamefont{J.~P.} \bibnamefont{Neirotti}},
  \bibnamefont{and} \bibinfo{author}{\bibfnamefont{S.}~\bibnamefont{Kais}},
  \bibinfo{journal}{J. Phys. Chem. A} \textbf{\bibinfo{volume}{102}},
  \bibinfo{pages}{9518} (\bibinfo{year}{1998}).

\bibitem[{\citenamefont{Sergeev and Kais}(1999)}]{Sergeev}
\bibinfo{author}{\bibfnamefont{A.~V.} \bibnamefont{Sergeev}} \bibnamefont{and}
  \bibinfo{author}{\bibfnamefont{S.}~\bibnamefont{Kais}},
  \bibinfo{journal}{Int. J. Quantum Chem.} \textbf{\bibinfo{volume}{75}},
  \bibinfo{pages}{533} (\bibinfo{year}{1999}).

\bibitem[{\citenamefont{Guevara and Turbiner}(2011)}]{PhysRevA.84.064501}
\bibinfo{author}{\bibfnamefont{N.~L.} \bibnamefont{Guevara}} \bibnamefont{and}
  \bibinfo{author}{\bibfnamefont{A.~V.} \bibnamefont{Turbiner}},
  \bibinfo{journal}{Phys. Rev. A} \textbf{\bibinfo{volume}{84}},
  \bibinfo{pages}{064501} (\bibinfo{year}{2011}).

\bibitem[{\citenamefont{Montgomery et~al.}(2018)\citenamefont{Montgomery, Sen,
  and Katriel}}]{PhysRevA.97.022503}
\bibinfo{author}{\bibfnamefont{H.~E.} \bibnamefont{Montgomery}},
  \bibinfo{author}{\bibfnamefont{K.~D.} \bibnamefont{Sen}}, \bibnamefont{and}
  \bibinfo{author}{\bibfnamefont{J.}~\bibnamefont{Katriel}},
  \bibinfo{journal}{Phys. Rev. A} \textbf{\bibinfo{volume}{97}},
  \bibinfo{pages}{022503} (\bibinfo{year}{2018}).

\bibitem[{\citenamefont{Jiao et~al.}(2022{\natexlab{a}})\citenamefont{Jiao,
  Zheng, Liu, Montgomery, and Ho}}]{PhysRevA.105.052806}
\bibinfo{author}{\bibfnamefont{L.~G.} \bibnamefont{Jiao}},
  \bibinfo{author}{\bibfnamefont{R.~Y.} \bibnamefont{Zheng}},
  \bibinfo{author}{\bibfnamefont{A.}~\bibnamefont{Liu}},
  \bibinfo{author}{\bibfnamefont{H.~E.} \bibnamefont{Montgomery}},
  \bibnamefont{and} \bibinfo{author}{\bibfnamefont{Y.~K.} \bibnamefont{Ho}},
  \bibinfo{journal}{Phys. Rev. A} \textbf{\bibinfo{volume}{105}},
  \bibinfo{pages}{052806} (\bibinfo{year}{2022}{\natexlab{a}}).

\bibitem[{\citenamefont{Turbiner et~al.}(2022)\citenamefont{Turbiner,
  Lopez~Vieyra, del Valle, and
  Julian~Nader}}]{https://doi.org/10.1002/qua.26879}
\bibinfo{author}{\bibfnamefont{A.~V.} \bibnamefont{Turbiner}},
  \bibinfo{author}{\bibfnamefont{J.~C.} \bibnamefont{Lopez~Vieyra}},
  \bibinfo{author}{\bibfnamefont{J.~C.} \bibnamefont{del Valle}},
  \bibnamefont{and}
  \bibinfo{author}{\bibfnamefont{D.}~\bibnamefont{Julian~Nader}},
  \bibinfo{journal}{Int. J. Quantum Chem.} \textbf{\bibinfo{volume}{122}},
  \bibinfo{pages}{e26879} (\bibinfo{year}{2022}).

\bibitem[{\citenamefont{Liverts and Barnea}(2011)}]{Liverts_2011}
\bibinfo{author}{\bibfnamefont{E.~Z.} \bibnamefont{Liverts}} \bibnamefont{and}
  \bibinfo{author}{\bibfnamefont{N.}~\bibnamefont{Barnea}},
  \bibinfo{journal}{J. Phys. A} \textbf{\bibinfo{volume}{44}},
  \bibinfo{pages}{375303} (\bibinfo{year}{2011}).

\bibitem[{\citenamefont{Fernández and Garcia}(2013)}]{FERNANDEZ2013580}
\bibinfo{author}{\bibfnamefont{F.~M.} \bibnamefont{Fernández}}
  \bibnamefont{and} \bibinfo{author}{\bibfnamefont{J.}~\bibnamefont{Garcia}},
  \bibinfo{journal}{Appl. Math. Comput.} \textbf{\bibinfo{volume}{220}},
  \bibinfo{pages}{580} (\bibinfo{year}{2013}).

\bibitem[{\citenamefont{Klaus and Simon}(1980)}]{KLAUS1980251}
\bibinfo{author}{\bibfnamefont{M.}~\bibnamefont{Klaus}} \bibnamefont{and}
  \bibinfo{author}{\bibfnamefont{B.}~\bibnamefont{Simon}},
  \bibinfo{journal}{Ann. Phys} \textbf{\bibinfo{volume}{130}},
  \bibinfo{pages}{251 } (\bibinfo{year}{1980}).

\bibitem[{\citenamefont{Efthimiou and Frye}(2014)}]{Harmonics}
\bibinfo{author}{\bibfnamefont{C.}~\bibnamefont{Efthimiou}} \bibnamefont{and}
  \bibinfo{author}{\bibfnamefont{C.}~\bibnamefont{Frye}},
  \emph{\bibinfo{title}{Spherical Harmonics in p Dimensions}}
  (\bibinfo{publisher}{WORLD SCIENTIFIC}, \bibinfo{year}{2014}).

\bibitem[{\citenamefont{Brownstein}(2000)}]{Brownstein}
\bibinfo{author}{\bibfnamefont{K.~R.} \bibnamefont{Brownstein}},
  \bibinfo{journal}{Amer. J. Phys.} \textbf{\bibinfo{volume}{68}},
  \bibinfo{pages}{160} (\bibinfo{year}{2000}).

\bibitem[{\citenamefont{Yang and de~Llano}(1989)}]{Yang}
\bibinfo{author}{\bibfnamefont{K.}~\bibnamefont{Yang}} \bibnamefont{and}
  \bibinfo{author}{\bibfnamefont{M.}~\bibnamefont{de~Llano}},
  \bibinfo{journal}{Amer. J. Phys.} \textbf{\bibinfo{volume}{57}},
  \bibinfo{pages}{85} (\bibinfo{year}{1989}).

\bibitem[{\citenamefont{Kocher}(1977)}]{Kocher1977}
\bibinfo{author}{\bibfnamefont{C.~A.} \bibnamefont{Kocher}},
  \bibinfo{journal}{Amer. J. Phys.} \textbf{\bibinfo{volume}{45}},
  \bibinfo{pages}{71} (\bibinfo{year}{1977}).

\bibitem[{\citenamefont{Baye}(2015)}]{BAYEPhysRep}
\bibinfo{author}{\bibfnamefont{D.}~\bibnamefont{Baye}}, \bibinfo{journal}{Phys.
  Rep.} \textbf{\bibinfo{volume}{565}}, \bibinfo{pages}{1 }
  (\bibinfo{year}{2015}).

\bibitem[{\citenamefont{Baye}(2011)}]{Baye_2011}
\bibinfo{author}{\bibfnamefont{D.}~\bibnamefont{Baye}}, \bibinfo{journal}{J.
  Phys. A} \textbf{\bibinfo{volume}{44}}, \bibinfo{pages}{395204}
  (\bibinfo{year}{2011}).

\bibitem[{\citenamefont{del Valle et~al.}(2023)\citenamefont{del Valle,
  Segura~Landa, and Nader}}]{delvalle2023two}
\bibinfo{author}{\bibfnamefont{J.~C.} \bibnamefont{del Valle}},
  \bibinfo{author}{\bibfnamefont{J.~A.} \bibnamefont{Segura~Landa}},
  \bibnamefont{and} \bibinfo{author}{\bibfnamefont{D.~J.} \bibnamefont{Nader}},
  \bibinfo{journal}{Phys. Rev. B} \textbf{\bibinfo{volume}{108}},
  \bibinfo{pages}{155421} (\bibinfo{year}{2023}).

\bibitem[{\citenamefont{del Valle and Turbiner}(2019)}]{radialQAHO}
\bibinfo{author}{\bibfnamefont{J.~C.} \bibnamefont{del Valle}}
  \bibnamefont{and} \bibinfo{author}{\bibfnamefont{A.~V.}
  \bibnamefont{Turbiner}}, \bibinfo{journal}{Int. J. Mod. Phys. A}
  \textbf{\bibinfo{volume}{34}}, \bibinfo{pages}{1950143}
  (\bibinfo{year}{2019}).

\bibitem[{\citenamefont{Turbiner et~al.}(2021)\citenamefont{Turbiner,
  Lopez~Vieyra, del Valle, and Nader}}]{TurbinerCompact}
\bibinfo{author}{\bibfnamefont{A.~V.} \bibnamefont{Turbiner}},
  \bibinfo{author}{\bibfnamefont{J.~C.} \bibnamefont{Lopez~Vieyra}},
  \bibinfo{author}{\bibfnamefont{J.~C.} \bibnamefont{del Valle}},
  \bibnamefont{and} \bibinfo{author}{\bibfnamefont{D.~J.} \bibnamefont{Nader}},
  \bibinfo{journal}{Int. J. Quantum Chem.} \textbf{\bibinfo{volume}{121}},
  \bibinfo{pages}{e26586} (\bibinfo{year}{2021}).

\bibitem[{\citenamefont{Turbiner and {Juan Carlos López
  Vieyra}}(2006)}]{TURBINER2006309}
\bibinfo{author}{\bibfnamefont{A.~V.} \bibnamefont{Turbiner}} \bibnamefont{and}
  \bibinfo{author}{\bibnamefont{{Juan Carlos López Vieyra}}},
  \bibinfo{journal}{Phys. Rep.} \textbf{\bibinfo{volume}{424}},
  \bibinfo{pages}{309} (\bibinfo{year}{2006}), ISSN \bibinfo{issn}{0370-1573}.

\bibitem[{\citenamefont{Contessi et~al.}(2017)\citenamefont{Contessi, Lovato,
  Pederiva, Roggero, Kirscher, and {van Kolck}}}]{CONTESSI2017839}
\bibinfo{author}{\bibfnamefont{L.}~\bibnamefont{Contessi}},
  \bibinfo{author}{\bibfnamefont{A.}~\bibnamefont{Lovato}},
  \bibinfo{author}{\bibfnamefont{F.}~\bibnamefont{Pederiva}},
  \bibinfo{author}{\bibfnamefont{A.}~\bibnamefont{Roggero}},
  \bibinfo{author}{\bibfnamefont{J.}~\bibnamefont{Kirscher}}, \bibnamefont{and}
  \bibinfo{author}{\bibfnamefont{U.}~\bibnamefont{{van Kolck}}},
  \bibinfo{journal}{Phys. Lett. B} \textbf{\bibinfo{volume}{772}},
  \bibinfo{pages}{839} (\bibinfo{year}{2017}).

\bibitem[{\citenamefont{Adams et~al.}(2021)\citenamefont{Adams, Carleo, Lovato,
  and Rocco}}]{Carleo2021}
\bibinfo{author}{\bibfnamefont{C.}~\bibnamefont{Adams}},
  \bibinfo{author}{\bibfnamefont{G.}~\bibnamefont{Carleo}},
  \bibinfo{author}{\bibfnamefont{A.}~\bibnamefont{Lovato}}, \bibnamefont{and}
  \bibinfo{author}{\bibfnamefont{N.}~\bibnamefont{Rocco}},
  \bibinfo{journal}{Phys. Rev. Lett.} \textbf{\bibinfo{volume}{127}},
  \bibinfo{pages}{022502} (\bibinfo{year}{2021}).

\bibitem[{\citenamefont{Kirscher et~al.}(2015)\citenamefont{Kirscher, Barnea,
  Gazit, Pederiva, and van Kolck}}]{Pederiva2015}
\bibinfo{author}{\bibfnamefont{J.}~\bibnamefont{Kirscher}},
  \bibinfo{author}{\bibfnamefont{N.}~\bibnamefont{Barnea}},
  \bibinfo{author}{\bibfnamefont{D.}~\bibnamefont{Gazit}},
  \bibinfo{author}{\bibfnamefont{F.}~\bibnamefont{Pederiva}}, \bibnamefont{and}
  \bibinfo{author}{\bibfnamefont{U.}~\bibnamefont{van Kolck}},
  \bibinfo{journal}{Phys. Rev. C} \textbf{\bibinfo{volume}{92}},
  \bibinfo{pages}{054002} (\bibinfo{year}{2015}).

\bibitem[{\citenamefont{Harris and Smith}(2005)}]{Harris}
\bibinfo{author}{\bibfnamefont{F.~E.} \bibnamefont{Harris}} \bibnamefont{and}
  \bibinfo{author}{\bibfnamefont{V.~H.} \bibnamefont{Smith}},
  \bibinfo{journal}{The Journal of Physical Chemistry A}
  \textbf{\bibinfo{volume}{109}}, \bibinfo{pages}{11413}
  (\bibinfo{year}{2005}).

\bibitem[{\citenamefont{Keeble and Rios}(2020)}]{KEEBLE2020135743}
\bibinfo{author}{\bibfnamefont{J.}~\bibnamefont{Keeble}} \bibnamefont{and}
  \bibinfo{author}{\bibfnamefont{A.}~\bibnamefont{Rios}},
  \bibinfo{journal}{Phys. Let. B.} \textbf{\bibinfo{volume}{809}},
  \bibinfo{pages}{135743} (\bibinfo{year}{2020}), ISSN
  \bibinfo{issn}{0370-2693}.

\bibitem[{\citenamefont{Sen et~al.}(2021)\citenamefont{Sen, Montgomery, Yu, and
  Katriel}}]{QUANTUMDOTS}
\bibinfo{author}{\bibfnamefont{K.~D.} \bibnamefont{Sen}},
  \bibinfo{author}{\bibfnamefont{H.~E.} \bibnamefont{Montgomery}},
  \bibinfo{author}{\bibfnamefont{B.}~\bibnamefont{Yu}}, \bibnamefont{and}
  \bibinfo{author}{\bibfnamefont{J.}~\bibnamefont{Katriel}},
  \bibinfo{journal}{The European Physical Journal D}
  \textbf{\bibinfo{volume}{75}} (\bibinfo{year}{2021}), ISSN
  \bibinfo{issn}{1434-6079},
  \urlprefix\url{https://doi.org/10.1140/epjd/s10053-021-00183-8}.

\bibitem[{\citenamefont{Kato}(1951)}]{Kato1951FundamentalPO}
\bibinfo{author}{\bibfnamefont{T.}~\bibnamefont{Kato}},
  \bibinfo{journal}{Transactions of the American Mathematical Society}
  \textbf{\bibinfo{volume}{70}}, \bibinfo{pages}{195} (\bibinfo{year}{1951}),
  \urlprefix\url{https://api.semanticscholar.org/CorpusID:53965384}.

\bibitem[{\citenamefont{Turbiner and Vieyra}(2016)}]{Zexpansion}
\bibinfo{author}{\bibfnamefont{A.}~\bibnamefont{Turbiner}} \bibnamefont{and}
  \bibinfo{author}{\bibfnamefont{J.~L.} \bibnamefont{Vieyra}},
  \bibinfo{journal}{Canadian Journal of Physics} \textbf{\bibinfo{volume}{94}},
  \bibinfo{pages}{249} (\bibinfo{year}{2016}),
  \eprint{https://doi.org/10.1139/cjp-2015-0366},
  \urlprefix\url{https://doi.org/10.1139/cjp-2015-0366}.

\bibitem[{\citenamefont{Jiao et~al.}(2022{\natexlab{b}})\citenamefont{Jiao, Xu,
  Zheng, Liu, Zhang, Montgomery, and Ho}}]{Jiao_2022}
\bibinfo{author}{\bibfnamefont{L.~G.} \bibnamefont{Jiao}},
  \bibinfo{author}{\bibfnamefont{L.}~\bibnamefont{Xu}},
  \bibinfo{author}{\bibfnamefont{R.~Y.} \bibnamefont{Zheng}},
  \bibinfo{author}{\bibfnamefont{A.}~\bibnamefont{Liu}},
  \bibinfo{author}{\bibfnamefont{Y.~Z.} \bibnamefont{Zhang}},
  \bibinfo{author}{\bibfnamefont{H.~E.} \bibnamefont{Montgomery}},
  \bibnamefont{and} \bibinfo{author}{\bibfnamefont{Y.~K.} \bibnamefont{Ho}},
  \bibinfo{journal}{J. Phys. B At. Mol. Opt. Phys.}
  \textbf{\bibinfo{volume}{55}}, \bibinfo{pages}{195001}
  (\bibinfo{year}{2022}{\natexlab{b}}).

\bibitem[{\citenamefont{del Valle}(0)}]{delValleLMM}
\bibinfo{author}{\bibfnamefont{J.~C.} \bibnamefont{del Valle}},
  \bibinfo{journal}{Int. J. Mod. Phys. C} \textbf{\bibinfo{volume}{0}},
  \bibinfo{pages}{2450011} (\bibinfo{year}{0}).

\end{thebibliography}

\end{document}